\documentclass[]{interact}
\usepackage{epstopdf}% To incorporate .eps illustrations using PDFLaTeX, etc.
\usepackage[caption=false]{subfig}% Support for small, `sub' figures and tables
\usepackage[natbibapa,nodoi]{apacite}
\setcitestyle{square}
\theoremstyle{plain}% Theorem-like structures provided by amsthm.sty

\theoremstyle{definition}

\theoremstyle{remark}

\begin{document}
% \articletype{ARTICLE TEMPLATE}% Specify the article type or omit as appropriate#
\title{AiR - An Augmented Reality Application for Visualizing Air Pollution}

\author{
\name{Noble Saji Mathews\textsuperscript{a,d}, 
Sridhar Chimalakonda\textsuperscript{b,d} and Suresh Jain\textsuperscript{c}}
\affil{Indian Institute of Technology, Tirupati, India\\ \textsuperscript{a}ch19b023@iittp.ac.in, Department of Chemical Engineering\\
\textsuperscript{b}ch@iittp.ac.in, Department of Computer Science and Engineering\\
\textsuperscript{c}sureshjain@iittp.ac.in, Department of Civil and Environmental Engineering\\
\textsuperscript{d}Research in Intelligent Software \& Human Analytics (RISHA) Lab,\\Department of Computer Science and Engineering, IIT Tirupati.}
% }
}

\maketitle

\begin{abstract}
Air quality is a term used to describe the concentration levels of various pollutants in the air we breathe. The air quality, which is degrading rapidly across the globe, has been a source of great concern. Across the globe, governments are taking various measures to reduce air pollution. Bringing awareness about environmental pollution among the public plays a major role in controlling air pollution, as the programs proposed by governments require the support of the public. Though information on air quality is present on multiple portals such as the Central Pollution Control Board (\textit{CPCB}), which provides Air Quality Index, that could be accessed by the public. However, such portals are scarcely visited by the general public. Visualizing air quality in the location where an individual resides could help in bringing awareness among the public. This visualization could be rendered using Augmented Reality techniques. Considering the widespread usage of Android based mobile devices in India, and the importance of air quality visualization, we present \textit{AiR}, as an Android based mobile application. \textit{AiR} considers the air quality measured by \textit{CPCB}, in a locality that is detected by the users' GPS or in a locality of users' choice, and visualizes various air pollutants  ($PM_1{}_0$, $PM_2{}_.{}_5$, NO$_2$, SO$_2$, CO, $O_3$ and NH$_3$) present in the locality and displays them in the users' surroundings. \textit{AiR} also creates awareness in an interactive manner about the different pollutants, sources, and their impacts on health .%This method allows the data to be easily understood by the public and inspire or aid further studies in other fields.

\end{abstract}

\begin{keywords}
Augmented reality; Visualizing pollution; Air pollution; Visual exploration; Covid-19 impact; Awareness
\end{keywords}

\section{Introduction}
\label{intro}

Urban air pollution is a serious problem across the world, which directly and indirectly affects the health of a large portion of the global population, especially in developing nations. It is ranked as one of the top ten causes of death by a recent study on the Global Burden of Disease (GBD) conducted by the World Health Organization (WHO) contributing to nearly 4.2 million premature deaths every year (WHO-2018. Ambient air pollution: Health impacts\footnote{\url{http://www.who.int/airpollution/ambient/health-impacts/en}},\cite{landrigan2018lancet}, GBD-2016). Burden of disease from ambient air pollution for 2016\footnote{\url{http://www.who.int/airpollution/data/AAP\_BoD\_results\_May2018\_final.pdf}}. Over the past few years, India has seen many of its cities ranked in the top 10 most polluted cities in the world.\footnote{\url{https://www.iqair.com/world-most-polluted-cities/world-air-quality-report-2019-en.pdf}} Frequent occurrences of thick haze in various parts of the country have become a major cause of concern. Despite considerable research and numerous measures taken by various agencies like CPCB,  MoEF\&CC in the country since 1981 \footnote{\url{https://cpcb.nic.in/displaypdf.php?id=aG9tZS9haXItcG9sbHV0aW9uL05vLTE0LTE5ODEucGRm}}, the ground reality is that the improvements are slow and often lack support from the stakeholders which in turn leads to poor enforcement \citep{wang2012air} Life expectancy of an average person can be cut short by illnesses \citep{chen2008systematic}, which may be caused and worsened by pollution.% A short-sighted view on air pollution justifies this suffering as something which has to be endured to gain some extra income.% 
Most of the highly polluted cities have very high population densities as well, however people with a short-sighted view endure the depreciation of air quality as a small price to pay for gaining a little more income \citep{jun2019air}.

%Air pollution kills about 2 million people every year in India alone.%
Air pollution has been reported as the third-highest cause of premature deaths in India ranking just above smoking, and over 1.2 million Indians have lost their lives due to exposure to unsafe air in 2017 alone \citep{polk2019state}. Common misconceptions such as ``cool fogs” during mornings which turn out to be a result of high concentrations of $PM_2{}_.{}_5$, fine particulate matter of 2.5 micrometers or less in diameter that have been found to be most harmful to human health, its small size allowing it to lodge deep into lungs and make its way into the bloodstream. 

%The issues in enforcement is due to failure in communication of the issues in a manner which is understandable to the general public.%
The issues in enforcement are commonly attributed to a failure in communication of the issues in a manner that is understandable to the general public \citep{kathuria2000industrial}. Realizing this issue, the Government of India together with IIT Kanpur, launched the National Air Quality Index (AQI) \footnote{\url{https://cpcb.nic.in/National-Air-Quality-Index}}to strengthen air quality information dissemination system and in turn to increase awareness. AQI has become the most commonly used criterion in many countries including India, to assess the ambient air quality \citep{bhaskar2010atmospheric}.

AQI has helped in quantifying air pollution, but it is also important to make it easy to understand and to visualize the impact of pollution \citep{dutta2009common}. Several studies have been performed to visualize air quality. Most of these approaches specify requirements of users to interact with dedicated handheld devices or web portals \citep{kim2010inair, devarakonda2013real, dutta2009common}. Also, these studies present graphical visualizations of air density, quality, and so on, in a geographical area. Providing users with interactive visualizations of air pollutants could greatly influence the actions taken towards improving air quality. Presenting these visualizations as mobile applications could ease the task of viewing air quality and can also be made as an integral part of the decision making of the public and government \citep{dutta2009common}. %However, to the best%
However, to the best of our knowledge, we are not aware of any other work that visualizes pollutant particles in the air through a mobile application and with the help of Augmented Reality, except for Aire, which was specifically developed for iOS platform \citep{torres2019aire}. Considering the extensive usage of the Android operating system in India\footnote{\url{https://gs.statcounter.com/os-market-share/mobile/india}}, providing air quality visualization as an Android based mobile application could be helpful for a wide population range in the country. The rapid development of mobile and web technologies have opened up many avenues for sharing data that had enabled developers in developing applications that are more user friendly and appealing to users. 

%Current means have not been effective at engaging the populace and bridging the gap in understanding of complex data.Simple plots, time series and scatter plots for various factors of air pollution are easy to create and hence are still in use despite mostly lacking the ability to express spatial relationships for lay person.\citep{li2016visualization}A spacio-temporal visualization model using the varied technologies at hand has hence become a subject of study in order to help improve public awareness about this topic, one that is suitable for air composition metrics and can be understood by the public as well as the government personnel so as to help make pollution control an integral part of every decision making procedure.

\section{Related Work}
\label{related work}
The technological advancements over the past decade has inspired many to develop a visualizable air quality metric based on national air quality standards. Some have been aimed at presenting the health impacts and others try to represent air quality data to the general public in ways that are easy to understand. Also, many of these were designed to assist in data interpretation and pollution control measures \citep{park2011visualization}.

A variety of attempts have been made in this field, many without relying on 3d visualizations as well. \cite{nurgazy2019cavisap} in their paper, present a context-aware system CAVisAP for outdoor air pollution visualization. The system provides context-aware visualization taking into account location and time. CAVisAP also checks users’ pollutant sensitivity levels and color vision impairments to provide personalized pollution maps.

\cite{huang2014smart} came up with Smart-MapReduce, which aims towards helping in visualizing urban pollution using cloud computing technology. Smart-MapReduce processes a large amount of 3D GIS data to make a 3D air pollution map. They aimed at speeding up the process compared to MPI and Hadoop's MapReduce. It enables using 3D spatial queries for analysis and 3D visualization. This has tackled the problems of displaying the vertical distribution of air pollution, which could not be displayed by common 2D air pollution maps.\cite{zahran20133d} used 3D volumetric clouds to display air pollution concentrations at different levels from the ground. The interface was used to obtain a 3D visualization of the future air quality impact caused by the proposed urban transport scheme. By utilizing this method to plot multiple parameters, they aimed at increasing the level of understanding of the pollution dispersion.

\cite{li2016visualization} have presented a study aimed to analyze variations of air pollutants during the period of one year in China, through visualizations. A heat map based on the pollutant values has been presented to facilitate easy understanding and analysis of changes in air quality. An interactive web-based application to enable users to view the distribution of air pollutants in China has also been developed. This provides a map which is embedded with visualizations of different pollutants across the country, and thus facilitates users in understanding air quality across China \citep{lu2017interactive}. Handheld devices to detect and report air quality have been developed with an aim to increase public awareness on air quality. These devices facilitate visualization of air quality by pairing with a mobile phone, which displays the distribution of pollutants on a map through a web interface \citep{dutta2009common}.

GvSig-3D has been developed to visualize the atmosphere in a geographical space by using spatial locality data provided by Google Earth. This data is used to render 3D models of buildings in the locality, onto which air temperature data is projected as colored the indices of air quality \citep{san20113d}. VRML-ISOSURFACES is another effort that was developed to help users in navigating through a geographical space, and consequently view the distributions of temperature and pollutants in the locality \citep{san20113d}.

% Augmented Reality has found itself being used in many different ways over the past few years. 
Augmented Reality is being widely used in various domains over the past few years. Education is one of the many fields that are being enhanced by developments in AR.\cite{dunleavy2014augmented} in their literature review point out AR to be an instructional approach looking for the context where it will be the most effective tool amongst the collection of strategies available to educators. \cite{hung2017applying} in their study shows how AR effectively enhances learning and how children are more enthusiastic when using AR [].
\cite{maier2009augmented} utilized AR to help students with learning chemistry. Their study showed that understanding of spatial structure not only enhances the understanding of chemistry but also has the potential to help scientists to understand if designed catalysts have desired spatial structures. 

%Augmented reality (AR) has also been used prior to this in a variety of forms. One of them is the work done by 
\cite{bronack2011role} have utilized AR to enhance learning of socio-scientific issues. AR has qualities of enhancing one's sense of presence invoke a feeling of immediacy, and promote immersion. As pointed out in the study by \cite{chang2013integrating}, the use of AR was well perceived by students and helped change their attitudes and inspire conversations on the topic of concern.

Some recent examples of projects undertaken to visualize air pollution are new York times interactive website \footnote{The NY website compares the users' city with the most polluted cities \raggedright\url{https://www.nytimes.com/interactive/2019/12/02/climate/air-pollution-compare-ar-ul.html}} and ``Aire", an app that was designed for iOS \citep{torres2019aire}. Such an app would be extremely useful, yet, to the best of our knowledge we know of no such app that focuses on air pollution data in the Indian context and which allows visualizing trends over time and at multiple locations while spreading awareness about different pollutants, their ill effects and sources that can be controlled. Also, currently there exists no app for visualization of air pollution through Augmented Reality on Android platform, motivating the need for our work. 

One limiting factor common to all of the approaches so far has been the inability to present air pollution as something of immediate concern to someone who is not directly experiencing its impacts or is unaware of them. Essentially, people fail to act when they do not care or when they do not know the implications of a particular aspect on them \citep{christiano2017stop}. A study by \cite{kollmussmind}, highlights several factors that exist as barriers to pro-environmental behavior. Unlike \cite{etde_6377513}, they do not attribute a direct relationship to environmental knowledge. Though it certainly acts as one of the factors, there are many other key concerns that they bring up in their study, such as lack of internal incentives, socio-economic factors, and old behavior patterns. 
Moreover, the reduction in air pollution itself could also have positive benefits in reducing preventable non-communicable diseases \citep{kjellstrom2010public}. These factors necessitates the development of solutions that could keep users well informed of the air quality, irrespective of the knowledge backgrounds. This could facilitate in bringing a sense of involvement, which can greatly help in reducing behaviors which are harmful to the environment. Most of the existing approaches in the literature, aimed towards visualizing air quality are either observed to display complex visualizations or are observed to be unavailable to a vast population in India. Hence, we present \textit{AiR} as an Android based mobile application capable of presenting air quality information in a concise format. Also, the approach of \textit{AiR} lets users draw their own conclusions rather than dumping a lot of information on them. This is something which we think may appeal to the millennial population.

\section{Design of AiR}
\label{design}
\begin{figure}
    \centering
    \includegraphics[width = \linewidth, height = 8cm]{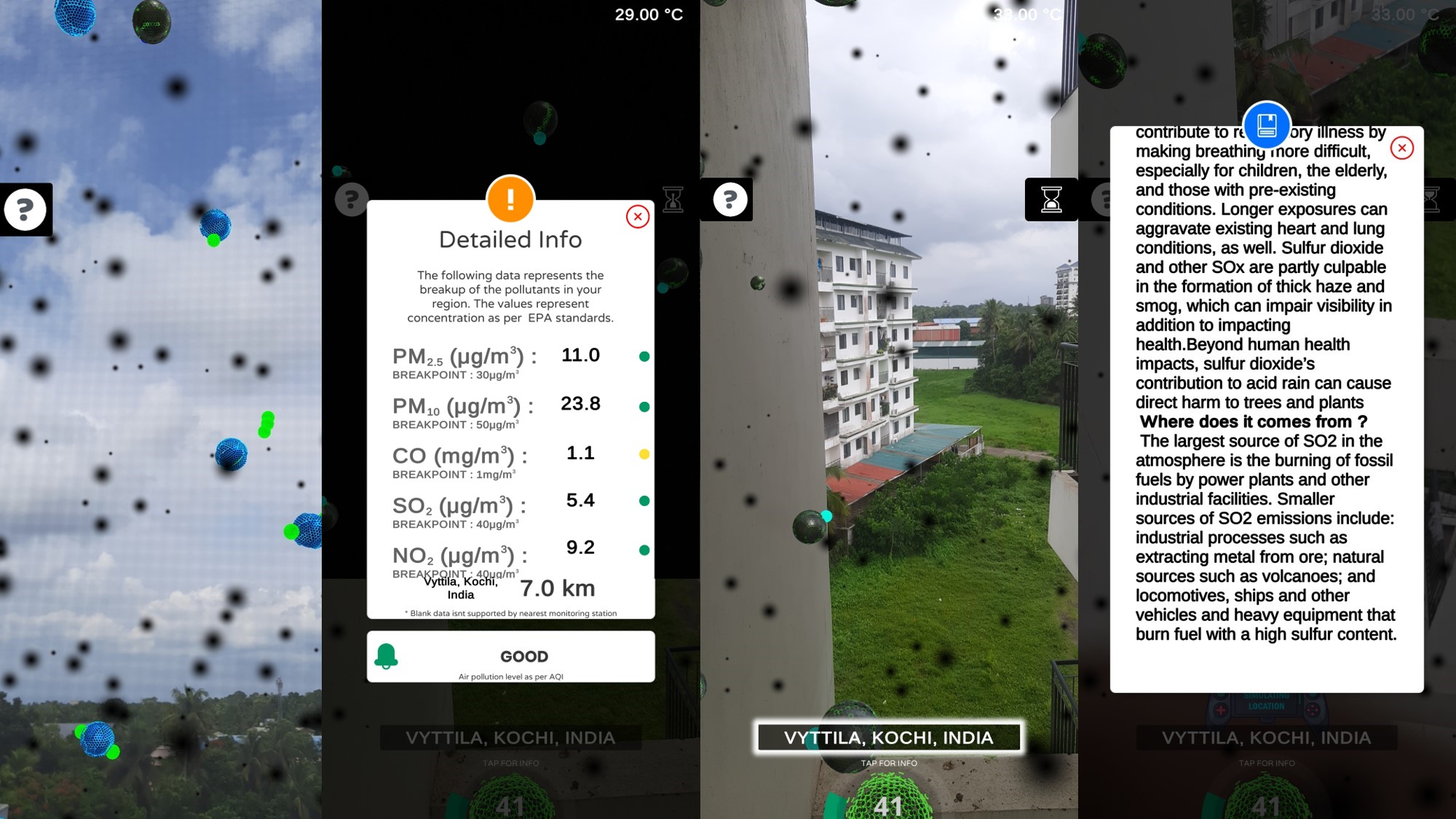}
    \caption{Snapshot of Live layout :\textit{AiR}}
    \label{fig:snapL}
\end{figure}
The technology for Augmented Reality (AR) is accessible through most of the commonly used smart phone's that support Google Play for AR Services\footnote{https://developers.google.com/ar/discover/supported-devices}. The rapid adoption of AR technology in various fields inspired us to pursue it as the basis of our study. AR takes advantage of virtual objects overlaid onto the surrounding real environment, resulting in a mixed reality where both coexist and thus provides a plethora of new possibilities. The current forms of graphical approaches have limited interactive and engaging multi-perspective visualizations of air pollution analysis. In this study, we focused on using mobile AR activity to build a system that can help a user easily grasp the quantitative information about the pollutants around and also visualize it. It also allows for a relative representation, which can be used to draw attention to the trends in the pollution data.

\textit{AiR}:Pollution Visualizer has been developed for making the user aware of the trends in air pollution over time by immersing them into an interactive AR simulated environment, which allows them to see and realize the effect of pollution in their cities and surroundings. Our work aims at overcoming the knowledge barriers to pro-environmental behavior \citep{kollmussmind}, thus effectively drawing the attention of citizens to air pollution. The underlying objective being, nudging the users to make responsible decisions and keep the atmosphere clean. The lockdown has confined people all over India within the four walls of their homes. With a lot of free time at their disposal, more and more people are turning to different forms of media for knowledge acquisition. So, we believe that the proposed application could be useful to the users, especially the millennials, to provide them with a way to visually understand the state of the environment in which they are living in. Once people look at the health benefits of living in a cleaner environment, they might try to inculcate positive habits, which when put into action, would help build up an environmentally conscious society. The trends followed by the various air pollution metrics can help people get a feel of where the environment is headed and help them realize the urgency of taking pro-environmental decisions, which can ensure a healthy future.

When the \textit{AiR} app is launched, the first screen which is shown to the user is the view as seen through the rear camera of the user's smartphone. The normal camera view is augmented by clickable objects which represents the various types of pollutants. It also has a dashboard which shows the real-time AQI value and color codes it according to the severity of the air pollution in that area. This first page queries the user’s location automatically and gets the GPS coordinates of the user. With the GPS coordinates, the app estimates the distance to the monitoring stations and finds the optimal monitoring station, which is located at the least distance.The latest air pollution data is retrieved from this station and used to create the visual representation of air quality. The data is analyzed using standards set by CPCB and the concentration of various pollutants obtained is used to calculate individual AQI values for the pollutants. The cumulative AQI score computed from these individual AQI values is displayed at the bottom of the screen for easy reference. In order to provide context to the numerical AQI value displayed on the screen, a radial gauge is displayed around the AQI monitor which is color coded to represent different air quality levels. The detailed view can be accessed by simply tapping on the AQI monitor, this view shows a breakup of all the pollutants their concentration and individual color scales for easy understanding. Primary breakpoints are also presented under each pollutant for reference. This view also shows the distance of the user from the closest live tower from which data has been sourced to account for ambient pollution metrics in use. 

The data in the detailed view is processed based on the AQI levels into relative counts to present a layout that makes it easy to compare the data from multiple locations and dates. On each run, an airspace of 10m x 10m x 5m ($500m^3$) around the user is used to visualize air pollution. The view is refreshed if the user reaches the boundaries of this airspace. The pollutants are randomly distributed into this airspace based on the air pollution data using Augmented Reality, which allows the user to experience this simulation and visualize the perspective through their own device and move around instilling a broader sense of involvement. The live view also retrieves and displays the current temperatures for an overall picture.

All the pollutants have unique 3D models to distinguish them in the simulated space. The molecules are built based on the chemical structures, and the atoms are color coded for consistency. The view is also built to be interactive, allowing the user to tap on any of the pollutants and learn about them. Discovering a pollutant and tapping on it reveals an info modal containing the details about the pollutant its sources and effects.
%The sources also include one's under the users control so that these may be kept in mind and remembered along with their ill effects on the public through the unique experience.%
The sources listed also include one's that an individual can help regulate by placing a check on such activities in addition to the industrial sources which are mentioned. Currently, there are 12 unique pollutants within the app [ $PM_1{}_0$, $PM_2{}_.{}_5$, NO, NO$_2$, $NO_x$, CO, SO$_2$,O$_3$,NH$_3$, C$_6$H$_6$, (CH$_3$)C$_6$H$_5$ and (CH$_3$)$_2$C$_6$H$_5$ ]. All of them have unique information panels within the app. The detailed info panel includes a pollution level indicator to signify the meaning of the current AQI level as per Indian standards to the end user.

\begin{figure}
    \centering
    \includegraphics[width = \linewidth, height = 8cm]{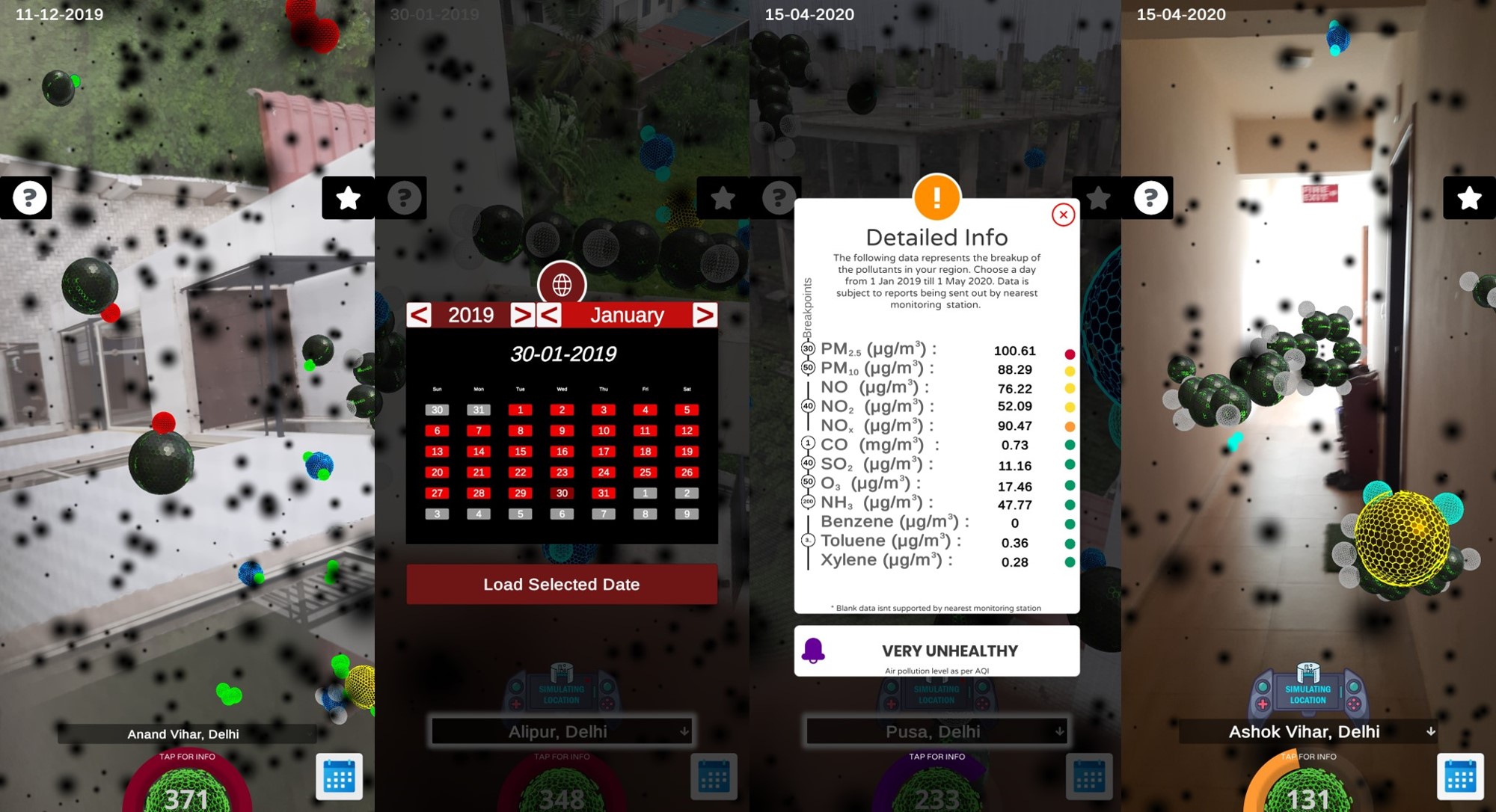}
    \caption{Snapshot of Historical layout: \textit{AiR}}
    \label{fig:snapH}
\end{figure}

On tapping the hourglass icon positioned on the right hand side on the screen, the user is taken to the historical layout which allows him/her to compare the data through AR from multiple locations from all over the country and on dates ranging from  the first of January 2019 to the first of May 2020. The data is stored on our server and is retrieved station wise. Currently, it serves 109 stations, but this is being regularly updated and will have all 230 stations soon. On load, the users GPS location is used to compare distance from a list of available stations using a optimised Haversines formula. The nearest location is queried and the data is updated into the calendar. The calendar can be opened with the button on the bottom right it shows the dates for which data is available for the selected station. On changing the selected date the view is refreshed and the new data is used to simulate the AR experience. The location used can be overridden using the dropdown menu to compare various locations across the country. When the user selects a location other than the closest one, a graphic indicator alerts the user that the location is being simulated to avoid confusion. Clicking on the hourglass icon turns into a star icon which can be clicked to return to the live view, data is refreshed and the view is updated on each iteration.

\section{Development}
\label{dev}

\textit{AiR}: Pollution Visualizer has been built as an AR based tool to improve the user's awareness about air pollution in India. It provides an easy to comprehend an overview of the state of the air quality of the locality in which the user resides and also gives an option to switch location to compare various locations and dates over the past year. The app was built with the help of the AR Foundation framework, which helps us to leverage the power of ARCore in Android. The workflow utilized to develop the app is shown in Figure 3.

\begin{figure}
    \centering
    \includegraphics[width = \linewidth]{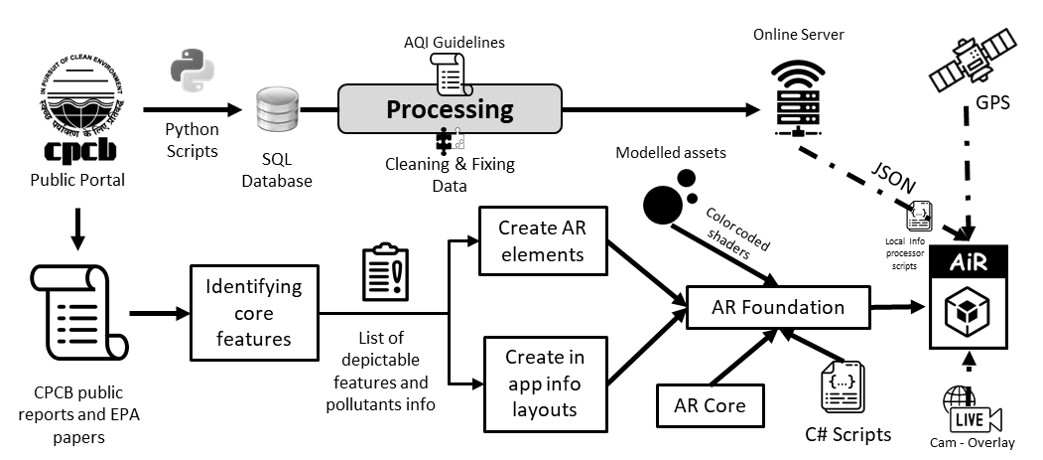}
    \caption{Workflow of building \textit{AiR}}
    \label{fig:dev}
\end{figure}

After deciding on the approach and identifying the features we wished to highlight, the first step was the collection of data. All raw data used in this project was sourced from the Central Pollution Control Board (CPCB) of India’s online portal. This portal provides us access to all the monitoring stations under CPCB and the various state-level control boards. Daily data since January 2019 till first of May 2020 was drawn using Python and the data was stored in a SQL database for further processing. This data was cleaned to remove invalid characters and to check for errors. After this the data was checked for missing dates for each of the stations. The dates with missing data were removed if they extended for periods longer than a 2-3 days after redrawing and confirming unavailability. Other missing  data was interpolated with the help of VBA scripts to match a linear function in order to avoid errors while working with the data.

After this, the next step was creating the models to be used for each of the pollutants. These were built as mesh groups with shaders applied to them so as to optimize performance. Each of the individual elements of the molecules were color coded for consistency throughout the app. They were given unique textures to make the designs look attractive and appeal to the user so that they would be curious to learn about all the different types of pollutants and learn about the same. The geometry used for each of the molecules was as per their molecular structures.

All the background scripts controlling the AR components were written in CSharp programming language. The quantitative control and spawning of the pollutants in the simulated airspace around the user was also done with CSharp Scripts. The national air quality index system adopted by India was utilized to get a relative scale of pollutants, which was then used for the visual representation. The breakpoints for each individual pollutant was obtained by calculating the respective subindexes by writing scripts for porting the system used by CPCB  itself into a format supported by mono scripts and IL2CPP to calculate AQI values. A max count of each pollutant was predefined after multiple tests on a range of devices to ensure good performance regardless of pollution levels and to maintain an upper cap so that the engine doesn't overload.

The fully processed data was supplied through a rest API as JSON responses for each location as per requirement in the app to minimize space requirements and background data used.  An optimised Haversian was also coded in to obtain the closest location to user each time the app is opened. On fresh launch, the app queries the server for historical data of the nearest monitoring station as well as the most recent data available. On overriding the location within the historical layout, the request is sent again for the new location. A script uses this data and populates the user interfaces based on the layout the user is on with available data. The individual AQI levels for each pollutant are processed on tapping on the AQI display at the bottom of the app to provide a layout containing numerical data such as concentrations and breakpoints but also individual color coding for easy understanding.

The range of the camera is also set to an optimal value of 3m in all directions (determined by evaluating the performance of the app on multiple devices) in order to encourage the user to move around and discover the particles. This was done to give the user a more interactive experience so that he/she does not just perceive it as a video. The script also displays the distance to the nearest tower, which is in use and a special indicator if the location is changed so as to prevent any ambiguity.

Another script is invoked once the data is processed in app as per query and the corresponding pollutants are spawned as per values calculated. It also checks if user is about to walk out of simulated area by keeping watch on distance from origin which is set on opening the app and calls the refresh method accordingly. The particles are spawned at random locations with small velocities and rotations so as to be dynamic and represent air as the fluid it is. Particulate matter which has no specific size but rather a range spawns at varied sizes within this range. User taps are also registered and raycasts are issued to identify particle being invoked. Data is displayed to the user for the corresponding particles with info about what it is, its molecular structure, how it is formed, and the ill effects it causes. This data is sourced from authorized Environmental Protection Agencies and public reports and warnings issued by CPCB.

\section{User Scenario}
\label{pic}

\begin{figure}
    \centering
      \begin{minipage}[t]{0.45\textwidth}
      \centering
    \includegraphics[width=0.6\textwidth, height = 8cm ]{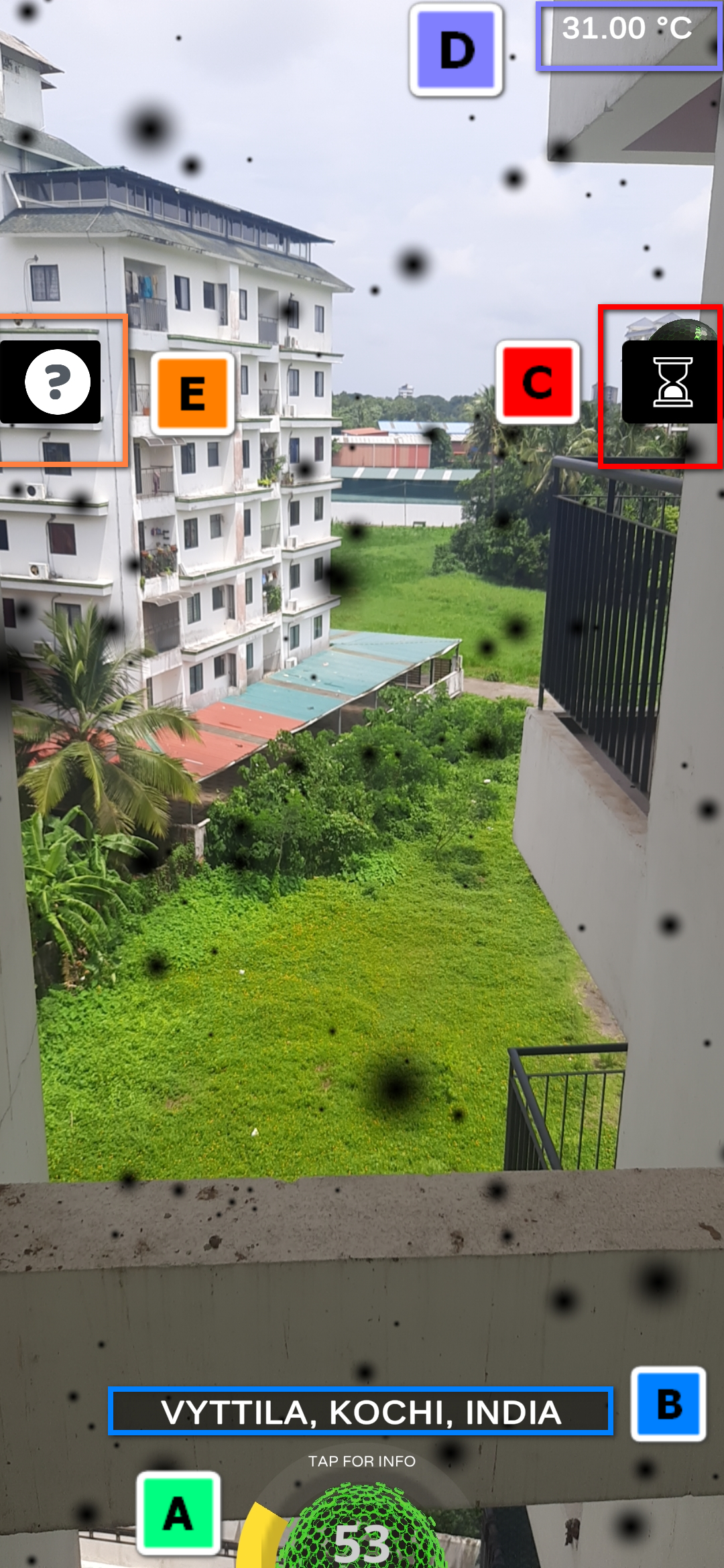}
    \caption{Start screen of \textit{AiR} depicting [A] AQI info-graphic button, [B]Current Location in use (Nearest) [C] Historical option toggle,[D] Temperature Info and [E] About option}
    \label{fig:us1}
      \end{minipage}
            \hspace{0.05\textwidth}
  \begin{minipage}[t]{0.45\textwidth}
  \centering
      \includegraphics[width=0.6\textwidth, height = 8cm ]{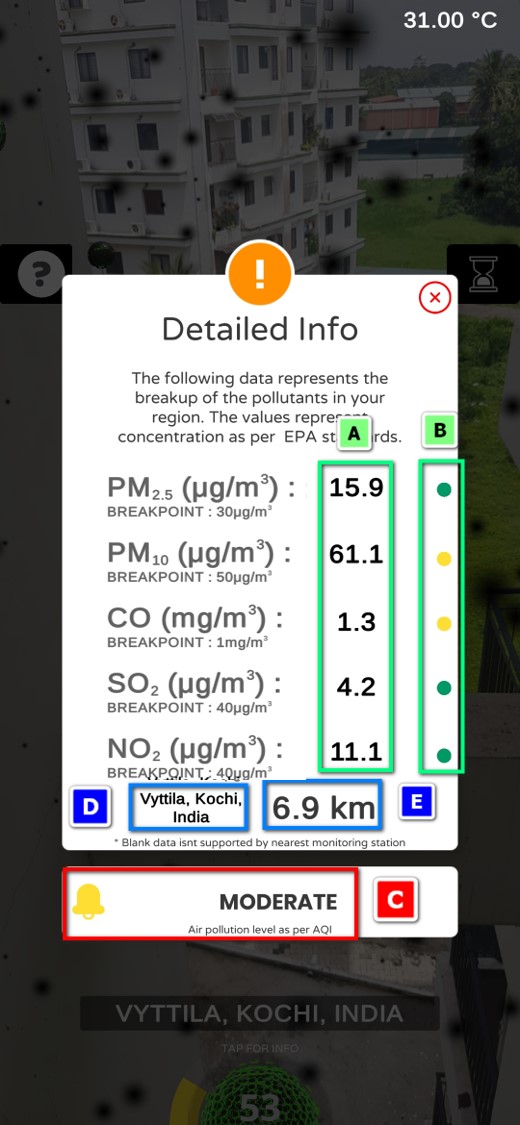}
    \caption{Snapshot of detailed info panel depicting [A] Pollutant Concentrations,[B] Individual Pollutant Concern Level , [C] Current Air Quality Level(both color coded), [D] Nearest station and [E] Distance from users current coordinates. }
    \label{fig:us2}
  \end{minipage}

\end{figure}

\begin{figure}
    \centering
          \begin{minipage}[t]{0.45\textwidth}
          \centering
    \includegraphics[width = 0.6\linewidth, height = 8cm ]{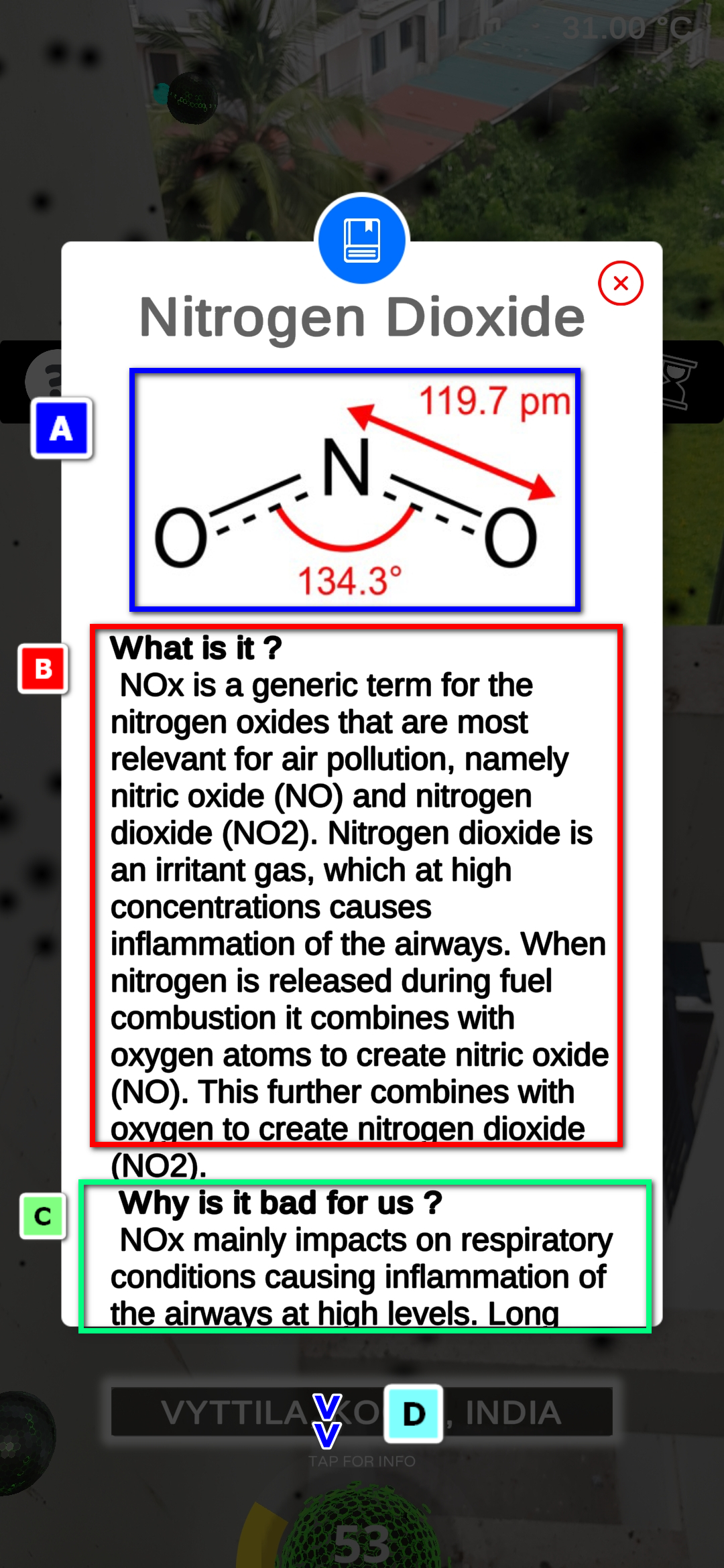}
    \caption{Snapshot of one of the pollutant info panels depicting [A] Molecular Structure,[B]Info about the pollutant,  [C] Ill effects due to pollutant and [D] Sources of pollutant controllable by user. }
    \label{fig:us3}
      \end{minipage}
            \hspace{0.05\textwidth}
  \begin{minipage}[t]{0.45\textwidth}
      \centering
    \includegraphics[width = 0.6\linewidth, height = 8cm ]{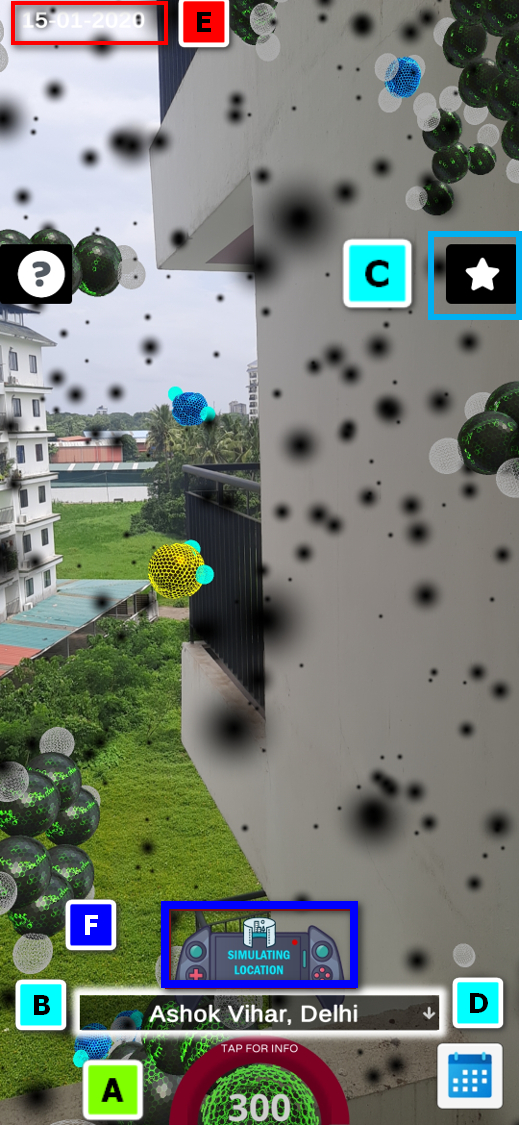}
    \caption{Historical layout \textit{AiR} depicting [A] AQI info-graphic button, [B]Location Dropdown [C] Live option toggle,[D] Date Selector [E] Selected Date Indicator and [F] Location override Indicator }
    \label{fig:us4}
  \end{minipage}

\end{figure}
\begin{figure}
    \centering
          \begin{minipage}[t]{0.45\textwidth}
	    \centering
    \includegraphics[width = 0.6\linewidth, height = 8cm ]{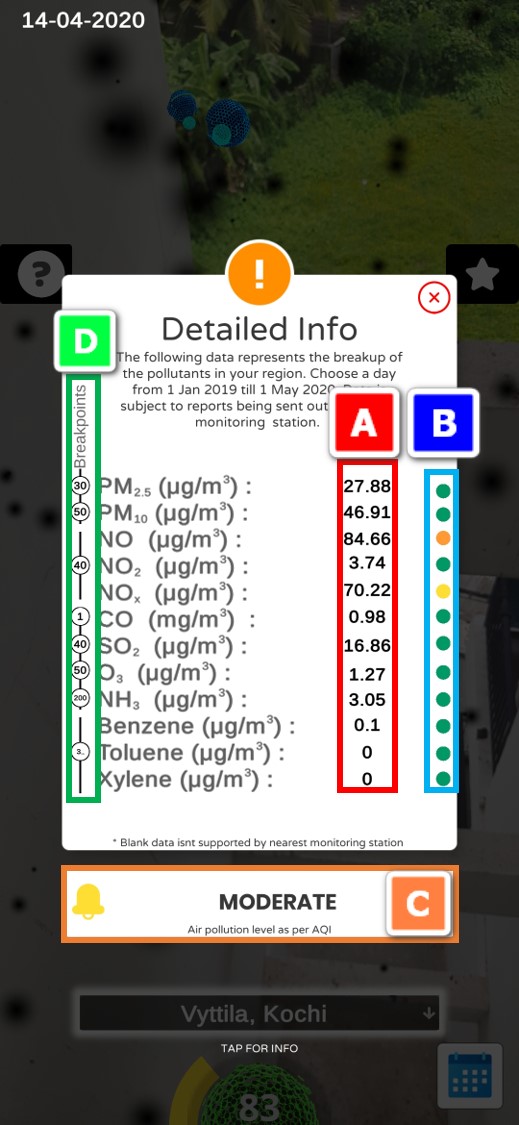}
    \caption{Snapshot of detailed info panel depicting [A] Pollutant Concentrations,[B] Individual Pollutant Concern Level , [C] Current Air Quality Level(both color coded) and [D]breakpoints for good level}
    \label{fig:us5}
      \end{minipage}
      \hspace{0.05\textwidth}
  \begin{minipage}[t]{0.45\textwidth}
      \centering
    \includegraphics[width = 0.6\linewidth, height = 8cm ]{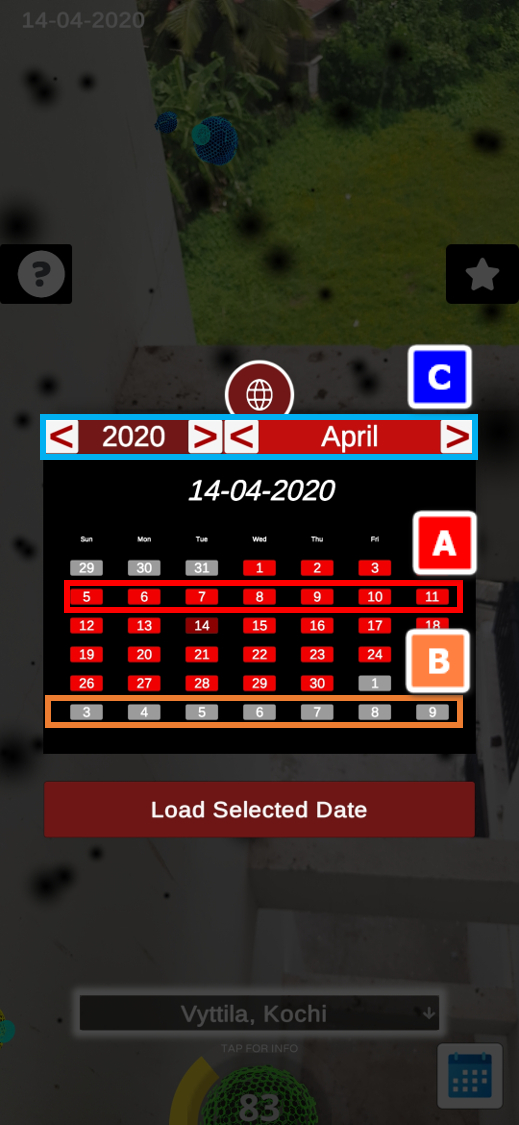}
    \caption{Date picker pop-up depicting [A] Available Dates
    [B] Unselectable/ Unavailable dates and [C] Toggles to 	  switch the month and year}
    \label{fig:us6}
  \end{minipage}
\end{figure}

Consider \textit{Moksha}, a teenaged girl who is suffering from asthma. Due to poor air quality during the day in her home state of Delhi, her parents have decided to take a transfer to a place with a better environment as per advice of the family doctor. Having arrived in her new locality in Kerala, she wants to keep tabs on the air quality before stepping out. She looks up on the Playstore and comes across \textit{AiR}, she had used a few apps before but was intrigued by the mention of the app using Augmented Reality which according to her was a buzzword in her peer group, time and also to be vary of safety measures to be taken when in outdoors. She downloads \textit{AiR} onto her Android mobile and starts the app. She is greeted to a view as shown in Figure \ref{fig:us1}. She walks around to discover several new particles floating around, some of which looked similar to molecules she had studied about in her science class. She taps on one of the particles and is shown a display with details about the pollutant, as shown in Figure \ref{fig:us2}. She reads through the info-panels to discover that there are many activities that cause this pollutant to be released into the atmosphere. Vowing to be more cautious about this next time, she decides to poke around the app a bit more. 

After reading the info about and discovering a few more particles, she gets an urge to somehow be able to compare her previous locality to this new one. She taps on the toggle to switch into historical view. By using the Calendar-like button, she loads data of multiple dates over the past months, which surprisingly were all in the green and at most yellow levels. \textit{Moksha} realizes she can actually change the location with the override button at the bottom of the screen, scrolling through the dropdown list she sees her previous district mentioned in it. On tapping it, an indicator comes up, showing her that she had changed the location from her nearest monitoring station to another one. Immediately the change was apparent the circular bar around the AQI meter had shot all the way around and the AQI value kept climbing past 300 in the monitor as shown in Figure \ref{fig:us4}. Walking around now she could feel the difference, the much higher density of pollutants and not to mention the many new types as well as number of them she could view just by moving her phone around.

\textit{Moksha} then taps on the info again this time she notices the colored indicators beside each pollutant and this made it easy for her to check multiple dates and actually make sense out of the breakup of pollutants. The app was an eye opener for her, even though she knew about the large amount of pollutants in the air she had never been able to get a feel of this until it would become so bad that there would be thick smog all across town making it hard to breathe. Jumping between many locations and back to his old locality reminded her of the talks her parents would have about the falling Air Quality. Moksha however, was not keen in participating in such conversations full of serious stuff but this, it made her feel as if she is actuality part of it, and it was not something she could brush aside especially considering her poor health. Looking into the info panel in the live view again, she could see the conclusion to be drawn from the current pollution levels. Since it was green and the tower was actually pretty close to her, it was definitely not as bad as her previous locality, which the app was showing warnings of being hazardous to all and not just sensitive groups on most days.

%Amazed by the simplicity and the way it had taught him about the pollutants and air quality in various parts of India he decides she should share this fun app with his friends so that they may also gain some additional knowledge about air pollution. She decides to join her friends to campaign for encouraging the rural villagers to use LPG gas cylinders rather than using firewood which she felt releases some of the pollutants which she saw in the app.

\section{Evaluation}
\label{Eval}
\textit{AiR} has been developed to educate the public about the air quality in India and inspire them to be part of the country's efforts to control air pollution since no country can succeed in achieving these goals without support of the public. On analyzing the data over the past few months during this period of lockdown we were able to see that in many cities pollution levels went down. Cities such as Delhi, which would top the charts of most polluted in the world, are slowly getting into the green levels. But the matter of concern is that in cities where the lockdown is being slowly lifted, the pollution levels are steadily going back to the same bad situation as before the lockdown was implemented.

Through \textit{AiR}, as an Android based interactive visualizer we wish to highlight this issue, there are also studies in progress that show that cities with bad air quality levels could harbor COVID-19 to a greater extent \citep{wu2020exposure}. The following plots show pollution levels since Jan 2020 (Before lockdown) till May 2020. The data here is shown till the end of May to show the concerning spikes in regions where lockdown was eased. Data for 3 cities Delhi, Patna and Ghaziabad which have come in the top 30 most polluted cities in the world.\footnote{\url{https://www.iqair.com/world-most-polluted-cities}}
\begin{figure}
    \centering
      \begin{minipage}[t]{0.45\textwidth}
      \centering
    \includegraphics[width=0.6\textwidth, height = 8cm ]{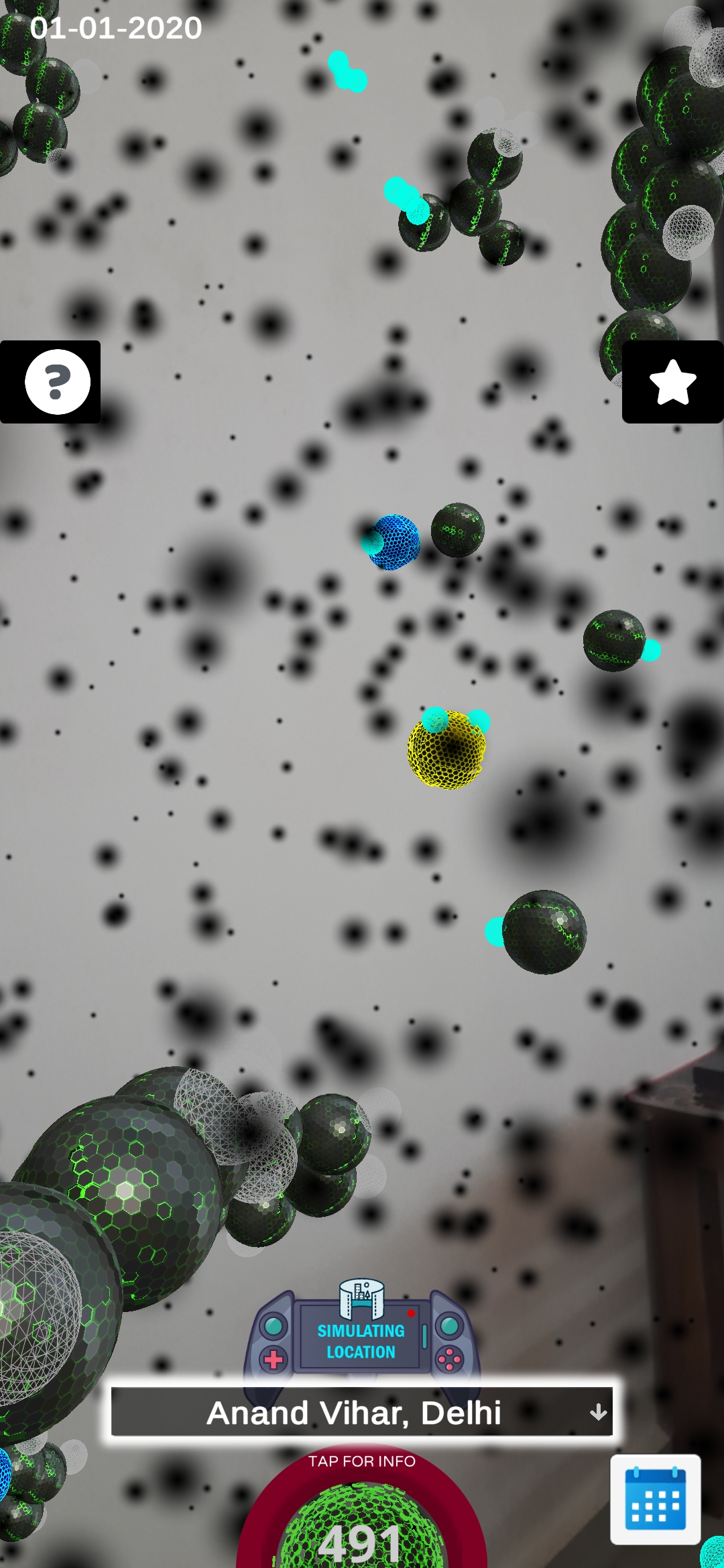}
    \caption{Depiction of Delhi before lockdown in Augmented Reality within \textit{AiR}}
    \label{fig:us7}
      \end{minipage}
            \hspace{0.05\textwidth}
  \begin{minipage}[t]{0.45\textwidth}
  \centering
      \includegraphics[width=0.6\textwidth, height = 8cm ]{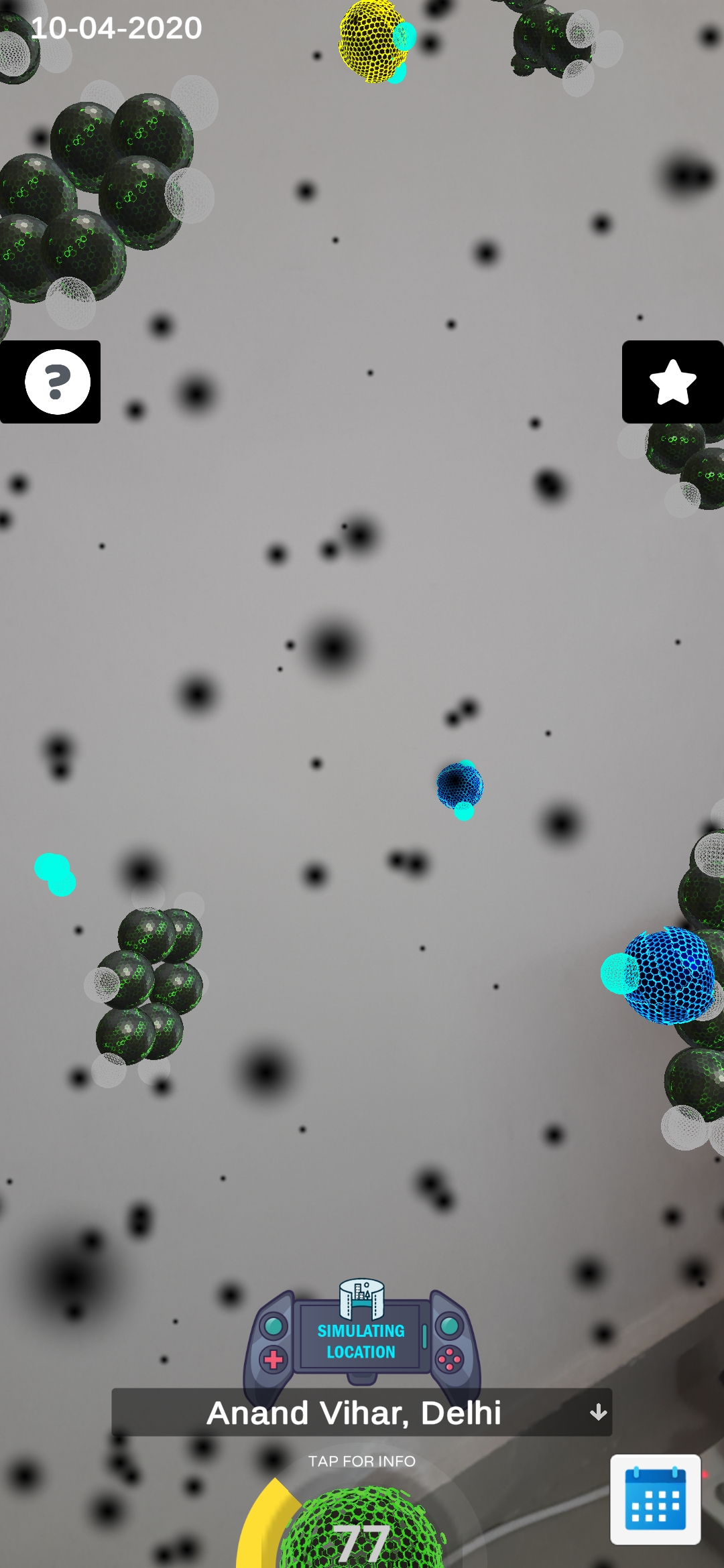}
    \caption{Depiction of Delhi after lockdown in Augmented Reality within \textit{AiR}}
    \label{fig:us8}
  \end{minipage}

\end{figure}

\begin{figure}
    \centering
    \includegraphics[width = \textwidth, height = 7cm]{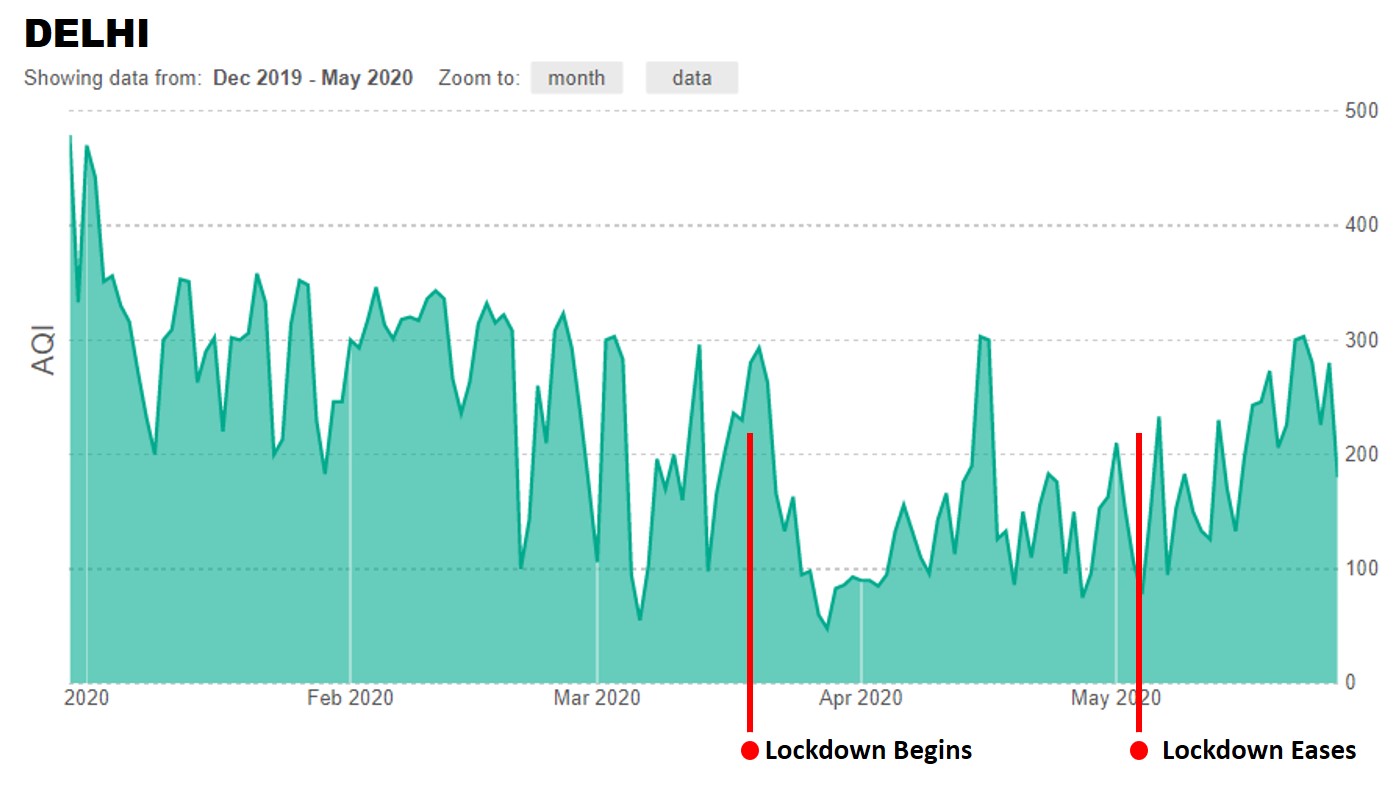}
    \caption{\textit{Delhi} AQI levels in 2020 showing a decrease in pollution due to lockdown but the concerning spike after slight relaxations}
\end{figure}
\begin{figure}
    \centering
    \includegraphics[width =\textwidth, height = 7cm]{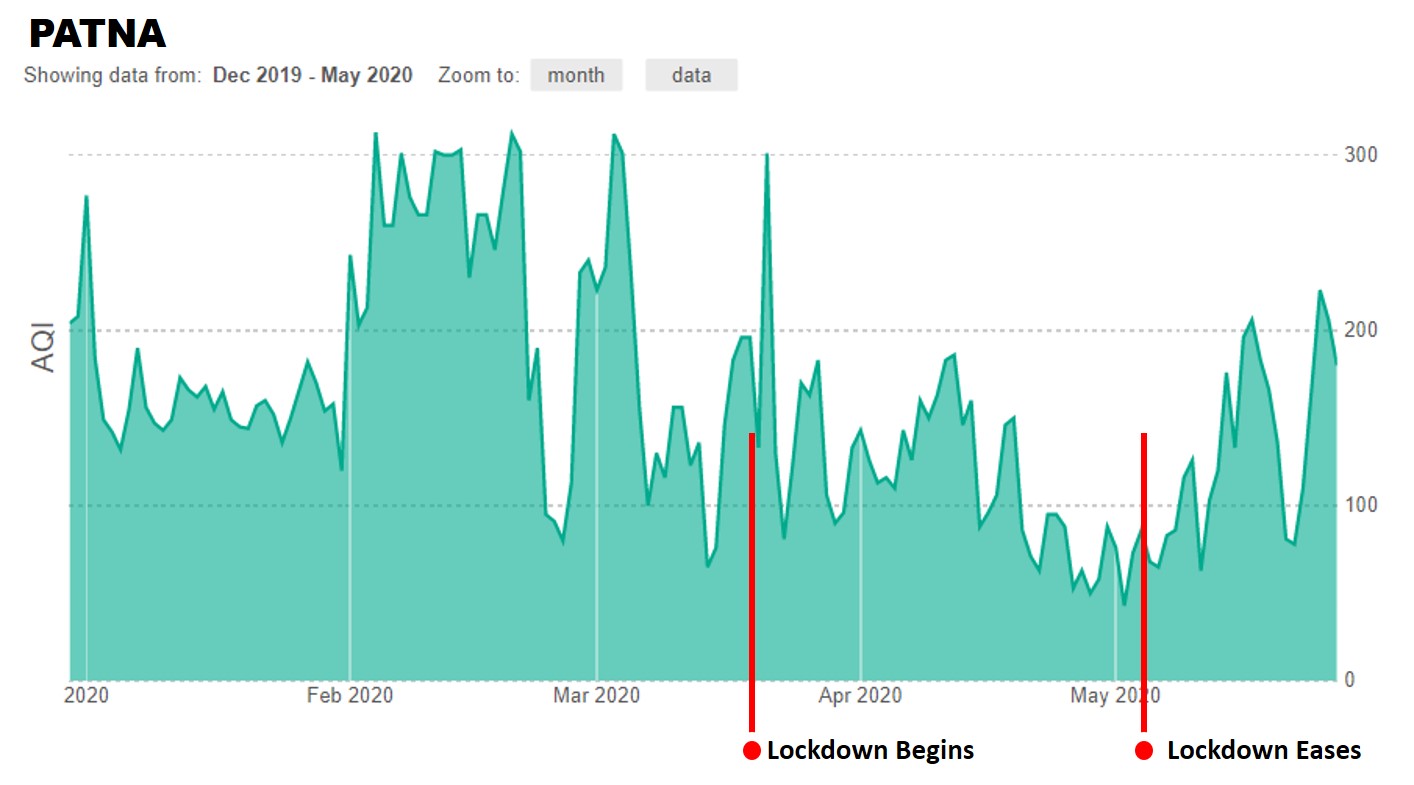}
    \caption{\textit{Patna} AQI levels in 2020 showing a decrease in pollution due to lockdown but the concerning spike after slight relaxations}
\end{figure}
\begin{figure}
    \centering
    \includegraphics[width = \textwidth, height = 7cm]{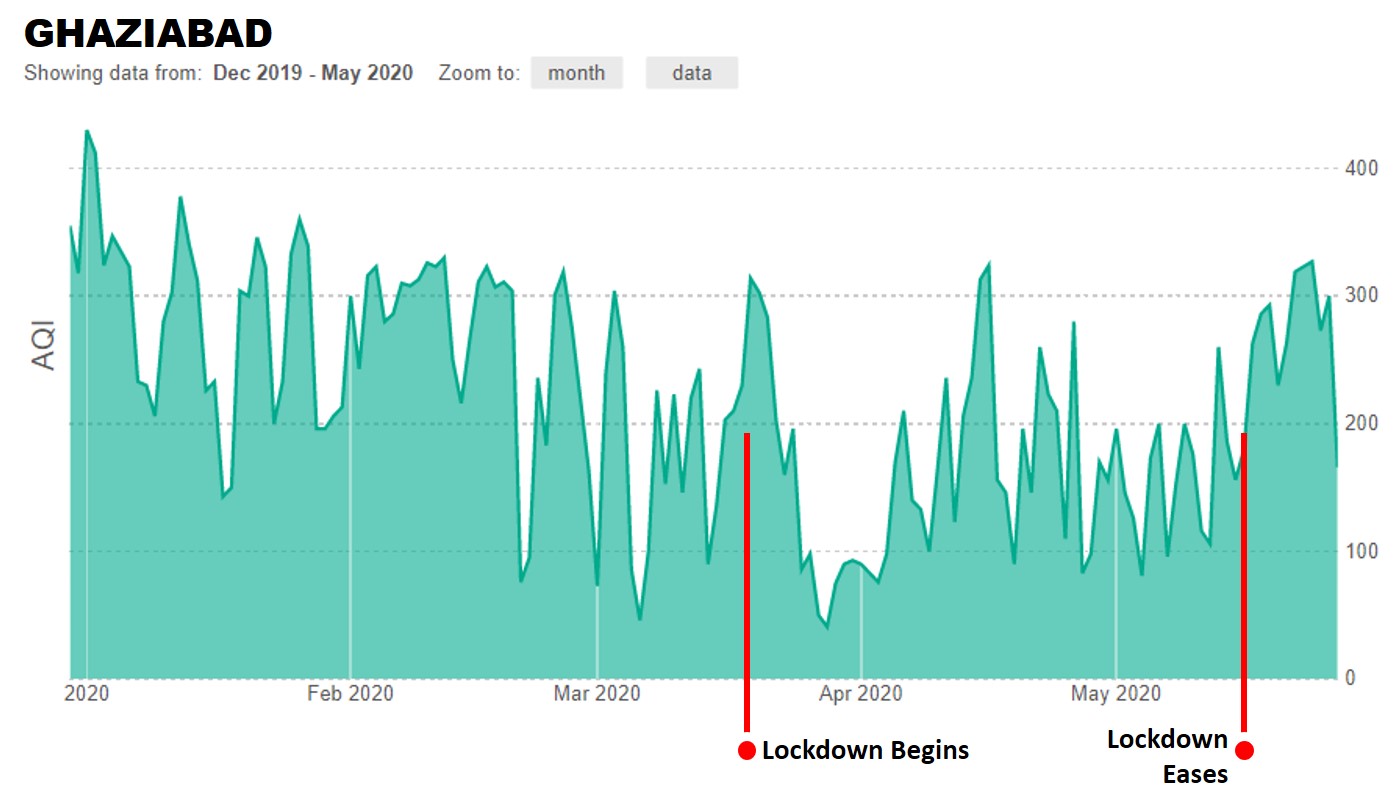}
    \caption{\textit{Ghaziabad} AQI levels in 2020 showing a decrease in pollution due to lockdown but the concerning spike after slight relaxations}
\end{figure}

\section{Results}
\label{results}

\begin{figure}
    \centering
    \includegraphics[width = \textwidth, height = 7cm]{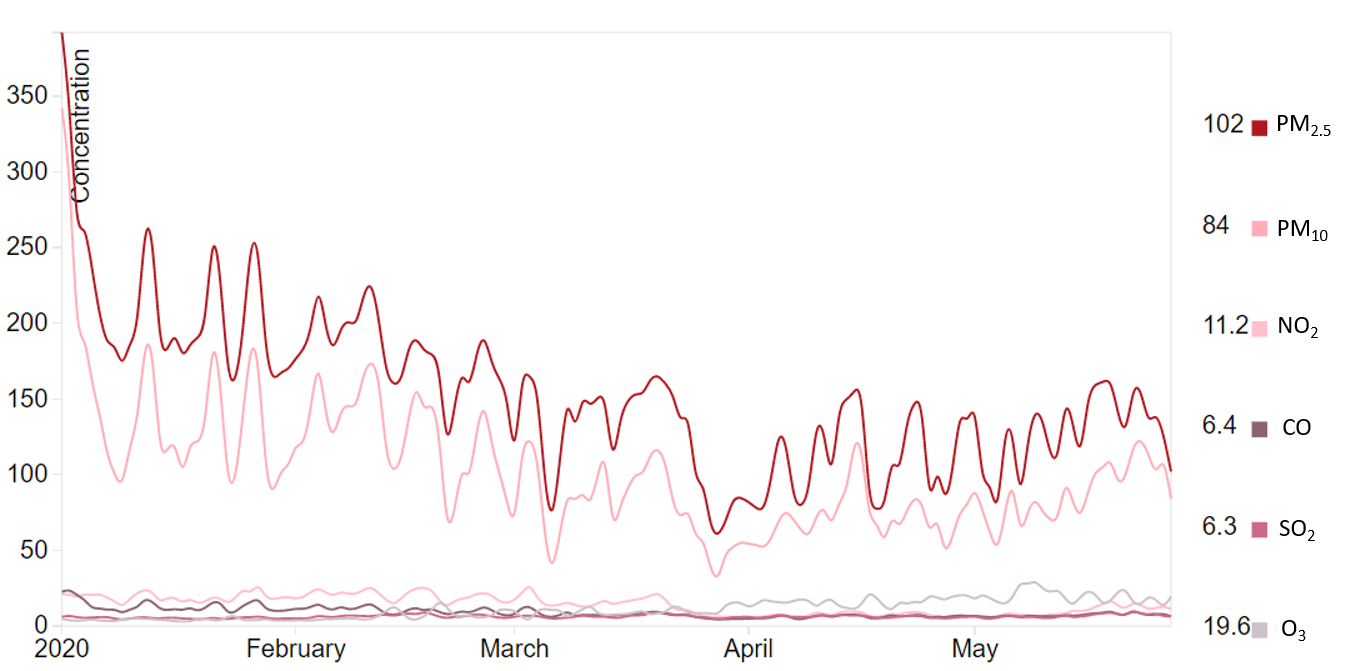}
    \caption{Individual Pollutant levels in Delhi since January}
\end{figure}

 The analyzed data indicated a clear dip in pollution levels overall during the lockdown, especially in places that were strongly affected. It also shows a concerning rise back to the initial pollutant levels at a much steeper rate. {fig:results}, depicts the trends of each of the pollutants in Delhi this year. It should also be noted that in India, pollution levels usually peak in the winter months due to the burning of crop residue in northern states. Also to note that 2020 has AQI values generally lower than last years, which is a good sign however, this needs not only to be maintained but also improved much more as a large share of Indian cities still are above the Harmful AQI levels.
 
  \textit{AiR} was built to address these concerns and to be a modern,interactive and hence appealing means to convey these concerns to the public. A Web portal was also built to make all the analysis available and enable access to all in a visual manner. \textit{AiR} is still continuously being updated to include more and more ways to visualize pollution concerns and instill an incentive by providing a sense of greater involvement and making the once intangible for many, tangible.

\section{Discussion and Limitations}
\label{limit}

%\textit{AiR} presents a method of visualizing air pollution in an interactive manner using rapidly developing technologies becoming available in everyday mobile devices that we spend a large share of our day on. 
With the advent of the internet and increasing affordability of devices to access it, the data which is available to a user is increasing day by day which may lead to information overload and at times may lead to important information getting ignored.\citep{savolainen2007filtering} \textit{AiR} is our attempt towards visualization of air pollution in an interactive manner, making it a memorable learning experience.
%True there are over 230 stations in India however it becomes pretty small when we look at the large areas it aims to cover and the population densities of those regions. This however becomes a issue only when trying to provide a personal report to individual users. The Air Quality Index represents the Outdoor Air Quality or Ambient Air Quality. This is not for Indoor Air Quality. If a Continuous Air Quality Monitoring Station is nearby the user it can influence the indoor air quality this has been pointed out in the about page. 
The Air Quality Index represents the Outdoor Air Quality or Ambient Air Quality. This however, should not be mistaken for Indoor Air Quality, if a Continuous Air Quality Monitoring Station is nearby, the user it can influence the indoor air quality. This has been pointed out in the about page. We currently have processed data for 109 stations across India since January 2019. We plan on adding data from the other stations, which where excluded for the initial phase of development due to inconsistencies in the data and the dates they started operation. Since the database is maintained separately from the app we can continue updating it over time.

The current version of \textit{AiR} is under active development, we are taking all feedback we receive to improve upon the systems in place. Since the app is still in closed testing as of writing this report user evaluation for the app has not yet been performed on a large enough scale, we plan on getting this done soon. We also wish to implement a street view-like system into the historical view when the location is changed from the current closest one. Currently the maximum number of particles is fixed, and then on basis of the data it is varied for a relative picture. We are working on making this dynamic as well as the addition of new data continuous without manual intervention. Also, the existing \textit{AiR} application can be used only on Android based mobile devices that are running on API-17 and should have Google AR services enabled for them. The future versions of the app could be made compatible to Apple devices and can also leverage web AR to become available on a large variety of devices.
%leverage web AR to become available on a large variety of devices.%

\section{Conclusion and Future Work}
\label{conclusion}

In this paper, we have discussed the importance of reducing air pollution levels and the need for bringing awareness among the public regarding air quality. Several visual methods have been employed in bringing awareness in multiple social and educational domains \citep{okada2001collaborative}. Also, there is an exponential increase in the use of mobile devices every month in India \citep{sheth2008chindia}, among which, majority of the public use Android devices. Hence, in this paper, we propose \textit{AiR} as an Android based mobile application that is capable of visualizing air quality in a specific locality. \textit{AiR} uses Augmented Reality to render visualizations as it helps users to interact and brings the virtual world closer to reality.

% motivation behind developing the \textit{AiR} app and the basic principles applied for development of the app. We are living in an age where most of the people are moving to mobiles devices and in India the growth of the number of mobile subscribers is increasing exponentially month on month so is the number of mobile phone consumers \citep{sheth2008chindia}. This is an ideal segment which can be targeted to make aware of the ill effects of air pollution and help them identify the activities which contributes to this pollution. Attempts to create virtual environment to provide environment education have been done \citep{okada2001collaborative} but augmented reality allows us to interact with it bringing it much closer to reality than the virtual world and gives a feel of a much more immersive experience to the user.
The following points are the main motivation towards designing \textit{AiR}:
\begin{itemize}
     \item \textit{Provide an interactive visualization experience of air pollutants through Augmented Reality} 
     %Create a fun and memorable experience for the user and give them something new to learn.}
     \item \textit{Help provide a means to bridge the barriers to Pro-Environmental behavior faced by the people.\citep{kollmussmind}}
 \end{itemize}

The current prototype version of \textit{AiR} could present information about the pollutants in the locality of users and might be helpful to the users in adopting better environment-friendly practices. 

% At this time we feel that the prototype  app which we built which help in giving people better information about the pollutants and in this way nudge more and more people to correct their ways and adopt more environment friendly practices.
We also presented an observation of pollution levels before and after COVID lockdown in India. We observe that COVID-19 pandemic and the subsequent stay at home orders had a great impact on the pollution levels in the atmosphere, which also adds to the existing research, which points to a lowering of pollutant levels \citep{mahato2020effect}. Through a graphical analysis of data processed in this study, we could conclude that there is a significant dip in the pollution levels after lockdown in India. But the steep rise in the air pollution levels, which were observed during this study, after easing of lockdown is a cause of concern, as the pollution levels might return back to the toxic levels in cities such as Delhi.

\label{Future Work}

The app currently runs on Android devices that have been granted access to ARCore services by Google. It can further be extended to support mobile devices running on other operating systems, and the system could be modified to work on a larger portion of the devices. Also in terms of the visualization, we plan on tweaking the parameters taken into account for the relative analysis to a more dynamic model. We plan to improve the AR experience by adding depth map support and improved occlusion that is promised with the upcoming releases of ARCore.  Also currently, not all stations are available for public view due to consistency issues in some of the newer stations and the stations that are no longer live. Once this data is fixed, we plan on updating \textit{AiR} to include about two times more locations that the users can choose from. We also plan on mapping several other related parameters that could be correlated and consequently observe the inherent patterns among the parameters which are otherwise not obvious.

% \section*{Acknowledgement(s)}
% %We would like to thank all the volunteers who tested the app for their valuable time and honest feedback.

\bibliographystyle{apacite}
\bibliography{AiR}

\end{document}